\documentclass{article}

\usepackage{arxiv}

\usepackage[utf8]{inputenc} 
\usepackage[T1]{fontenc}    
\usepackage{hyperref}       
\usepackage{url}            
\usepackage{booktabs}       
\usepackage{amsmath,amsfonts,amssymb}       
\usepackage{nicefrac}       
\usepackage{microtype}      
\usepackage{cleveref}       
\usepackage[noend]{algorithmic}
\usepackage{algorithm}
\usepackage{tikz}
\usepackage[]{xcolor}
\usepackage{pgfplots}
\pgfplotsset{compat=1.18}
\usepackage{graphicx}
\definecolor{changed}{RGB}{204,0,0}
\definecolor{unchanged}{RGB}{211,215,207}
\definecolor{TP}{RGB}{138,226,52}
\definecolor{TN}{RGB}{88,175,245}
\definecolor{FP}{RGB}{239,41,41}
\definecolor{FN}{RGB}{252,175,62}
\usepackage{natbib}
\usepackage{doi}
\usepackage{multirow}

\title{Deep Unsupervised Learning for 3D ALS Point Cloud Change Detection}

\date{}

\usepackage{authblk}

\author[1,2\thanks{\tt{iris.de-gelis@irisa.fr}}      \thanks{Iris de~G\'{e}lis is also with the Chair of Data Science in Earth Observation as a Beyond Fellow in the International Future Lab AI4EO, Technical University of Munich (TUM), Germany. Her work is also partly funded by the CNES, Toulouse, France.}]{Iris de~G\'{e}lis}
\author[3]{Sudipan Saha}
\author[4]{Muhammad Shahzad}
\author[5]{Thomas Corpetti}
\author[2]{S\'{e}bastien Lef\`{e}vre}
\author[4]{Xiao Xiang Zhu}

\affil[1]{Magellium, Toulouse, France}
\affil[2]{IRISA, UMR 6074, Universit\'{e} Bretagne Sud, Vannes, France}
\affil[3]{Yardi School of Artificial Intelligence, Indian Institute of Technology Delhi, New Delhi, India}
\affil[4]{Technical University of Munich (TUM), Chair of Data Science in Earth Observation (SiPEO), Munich, Germany}
\affil[5]{CNRS, LETG, UMR 6554, Rennes, France}
\renewcommand{\shorttitle}{Deep unsupervised 3D point cloud change detection}

\hypersetup{
pdftitle={Deep Unsupervised Learning for 3D ALS Point Clouds Change Detection},
pdfauthor={Iris de~G\'{e}lis, Sudipan Saha, Muhammad Shahzad, Thomas Corpetti, S\'{e}bastien Lef\`{e}vre, Xiao Xiang Zhu},
pdfkeywords={3D point clouds, change detection, self-supervised learning, unsupervised deep learning, Aerial LiDAR Survey},
}

\begin{document}
\maketitle

\begin{abstract}
	Change detection from traditional 2D optical images has limited capability to model the changes in the height or shape of objects. Change detection using 3D point cloud from photogrammetry or LiDAR surveying can fill this gap by providing critical depth information. While most existing machine learning based 3D point cloud change detection methods are supervised, they severely depend on the availability of annotated training data, which is in practice a critical point. To circumnavigate this dependence, we propose an unsupervised 3D point cloud change detection method mainly based on self-supervised learning using deep clustering and contrastive learning. The proposed method also relies on an adaptation of deep change vector analysis to 3D point cloud via nearest point comparison. Experiments conducted on an aerial LiDAR survey dataset show that the proposed method obtains higher performance in comparison to the traditional unsupervised methods, with a gain of about 9\% in mean accuracy (to reach more than 85\%). Thus, it appears to be a relevant choice in scenario where prior knowledge (labels) is not ensured. The code will be made available at \url{https://github.com/IdeGelis/torch-points3d-SSL-DCVA}.
\end{abstract}


\section{Introduction}\label{sectionIntroduction}
Considering the rapid evolution of our landscapes, change detection from bi-temporal satellite/aerial 2D images is one of the most important applications of remote sensing \citep{shi2020change, LI2022102926} and earth observation (map updates, damage identification, etc.). Currently, most existing change detection methods employ optical or Synthetic Aperture Radar (SAR) data \citep{saha2019unsupervised,saha2020building}. Both Siamese network-based supervised methods \citep{zhan2017change} and deep transfer learning-based unsupervised methods \cite{saha2019unsupervised} have been proposed in this context. However, optical images have limited capability to model the changes in the shape of the object, e.g., the change of building height caused by construction work. While SAR images are more suitable for such applications \citep{saha2020building}, they are not visually salient and analyzing them requires advanced domain expertise.  Thus, environment understanding from optical and SAR data has a significant shortfall in  tasks where  depth information is critical. 
\par
An interesting alternative consists in using 3D Point Clouds (PCs) since such data fill the above-mentioned gap and have become more accessible recently. These PCs can be obtained using either photogrammetric reconstruction or Light Detection and Ranging (LiDAR) acquisition. In particular, LiDAR sensor through aerial LiDAR surveying (ALS) allows to obtain 3D PCs at large territory scale.  While 3D point clouds change detection is popular \citep{qin20163d,shirowzhan2019comparative, degelis2021change}, most of the existing methods are based on a rather traditional processing approach. In particular, a vast majority of methods relies on thresholding (e.g., with Otsu algorithm) the difference of Digital Surface Models (DSM) obtained through a prior rasterization of PCs. The change maps are sometimes further refined by morphological operation to filter out false detections \citep{stal2013airborne, dos2021use}. 
However, the rasterization process implies a drastic loss of information. To counter this, some other methods proposed to directly process the 3D PCs through thresholding the 3D distances \citep{girardeau2005change,lague2013accurate, liu20213d} or more refined rule-based approaches often combining both distance and geometric information \citep{awrangjeb2015building, xu2015detection,dai2020object}. Although unsupervised, the rule-based methods are multi-step procedures likely to propagate errors from steps to steps. Thus, \citet{tran2018integrated} proposed a single-step supervised machine learning method to get rid of multi-step errors. Nevertheless, this traditional machine learning method still requires a feature extraction preliminary step. 
In the last decades, deep learning provided convincing results in remote sensing with a single network able to extract and select features, and further classify data depending on the application. However, 
deep learning remains rarely used in this context, 
calling for more development. 
Indeed, it has been demonstrated that deep learning outperforms both machine learning and traditional methods in numerous remote sensing applications, and compared to rule-based methods, it is less specific to one particular dataset. So far, only one supervised method has been proposed for change classification 
at point level \citep{degelis2023siamese}. However, supervised 3D PC change detection methods require training data with annotation statistically similar to the test data \citep{degelis2021change}. The tedious ground truth 
labelling step limits its adaptation to new applications and geographic areas. While some unsupervised learning methods have been proposed for 2D change detection \citep{saha2019unsupervised} (e.g., for optical and SAR data), their adaptation for PCs is not straightforward due to the particular characteristics of 3D PCs, e.g., sparsity of data and lack of point-to-point correspondence between pre-change and post-change PCs, surely explaining the lack of unsupervised learning method in the literature \citep{xiao20233d}. Thereby, in this paper we decide to explore unsupervised deep learning for binary change detection through processing directly the raw 3D PCs without any rasterization. 

\par
Self-supervised learning has recently emerged as a popular topic in computer vision and remote sensing \citep{stojnic2021self, wang2022ssl}. It consists in finding strategies to make the network training by itself without the use of external annotated ground truth. Thus, self-supervised learning belongs to the more general class of unsupervised learning. It has also been explored in the context of bi-temporal remote sensing analysis \citep{saha2022self} and multi-modal learning in remote sensing \citep{Heidler2023}.
  Motivated by this, we propose a self-supervised learning mechanism to train a network from unlabeled bi-temporal PC data. Following this, we employ this network as a feature extractor and modify Deep Change Vector Analysis (DCVA) \citep{saha2019unsupervised} using a nearest neighborhood comparison of PCs to account for the absence of point-to-point correspondence between pre-change and post-change 3D points. 
\par
The contributions of this work are as follows:
\begin{enumerate}
\item A deep learning unsupervised change detection method for remote sensing 3D PC data. To the best of our knowledge, this is the first one that relies on neural networks;
\item A training strategy for deep feature extractor based on self-supervised learning paradigm;
\item An adaptation of DCVA to raw 3D point clouds for bi-temporal deep features comparison;
\item Experimental comparisons to the state-of-the-art, i.e., existing supervised and unsupervised approaches, as well as an application of DCVA on deep features extracted by a pre-trained network on a public dataset for semantic segmentation (transfer learning scenario, as done in \citep{saha2019unsupervised} in the 2D case).
\end{enumerate}
This paper is organized as follows. We review related works in Section \ref{sectionrelwork}.
The proposed method is detailed in Section \ref{sectionProposedMethod} while experimental validation is presented in Section \ref{sectionExperiments}. Finally, a conclusion is proposed in Section \ref{sectionConclusion}.

\section{Related work}\label{sectionrelwork}

Though several methods have been proposed to perform change detection into 3D PCs, a large majority of them still relies on conventional processing approaches:  comparison of rasterized version of PCs \citep{xu2015using, okyay2019airborne, ZOVATHI2022102767,CSEREP2023103174}, distance computation \citep{girardeau2005change,lague2013accurate, liu20213d}, scene object extraction and bi-temporal comparison \citep{awrangjeb2015building,siddiqui2017novel,dai2020object} or machine learning with hand-crafted features \citep{tran2018integrated}. Even if deep learning is now well established in remote sensing \citep{zhu2017},  it is still a new field of research when dealing with change extraction in 3D data. While \cite{XIU2023103150} propose to detect damaged building after an earthquake using a supervised deep network on raw 3D point clouds, their method is not a suitable change detection method in the sense that collapsed buildings are only recovered from a point cloud acquired at a single date by looking at the shape of the building. Therefore, to the best of our knowledge, we counted only three studies relied on deep networks to highlight changes into bi-temporal 3D data. Among them, \cite{zhang2019detecting} propose to apply feed forward and Siamese convolutional neural networks on 2.5D rasterization of PCs into Digital Surface Models (DSMs). The results consist in a binary change label (change or no change) for each 2D patch. In \cite{KU2021192}, the authors experiment a Siamese architecture based on graph convolution to deal with raw 3D PCs. However, results  still remain at the scene/patch level and not at the point level. Finally, \cite{degelis2023siamese} propose the Siamese KPConv network to perform multi-change segmentation at the point level, in a supervised way. These three fully-supervised methods imply the availability of a fully annotated dataset, which can be challenging in some application contexts. Indeed, to obtain accurate labelization of changes at the point level, a manual annotation of each point of PCs is required, which is very time consuming. Therefore, unsupervised approaches constitute an interesting alternative. However, to the best of our knowledge, there are no unsupervised deep learning-based methods to tackle change detection between PCs.

Conversely to 3D PCs, change detection in 2D remote sensing images has been far more studied using deep learning based methods since the processing of 2D images is simplified by the regularity of the pixel grid which makes convolutions easier. To tackle the aforementioned difficulties related to data annotation, numerous deep frameworks have been designed to address this unsupervised change detection task. As highlighted in \cite{shi2020change}, a first category of methods is based on the \textit{generation of credible change pseudo-labels} to train a deep model. Generation of training data relies on various ideas such as combination of unsupervised traditional methods (e.g., change vector analysis (CVA)) \citep{song2018change, li2021unsupervised, seydi2021new, FANG2022102749}, fuzzy clustering \citep{gao2016automatic,zhan2018log, zhang2021robust}, metric learning \citep{zhao2019incorporating} or even unsupervised deep framework, e.g., auto-encoders (AEs) \citep{gong2017feature} or generative adversarial networks (GANs) combined with metric learning \citep{tang2021unsupervised}. 
To counter the class imbalance problem (i.e., changed areas are in minority compared to unchanged ones), additional GAN can be used to enrich changed class pseudo-labels \citep{zhang2021robust}. In general pixels are classified into three categories: changed, unchanged and uncertain. Only certain pixels are taken into account for the loss computation. Even if it has been shown that final change maps predicted by the deep network are more accurate than the original pseudo-change classification, these methods may somehow be limited by the pseudo-label quality. 

Therefore, a second category of methods is based on \textit{latent change map generated by deep features}. Transfer learning is a common strategy to train the deep model to extract useful features \citep{saha2019unsupervised}. However, transfer learning still requires the availability of an annotated source dataset. Therefore, fully unsupervised networks such as AEs \citep{lv2018deep, bergamasco2019unsupervised,kalinicheva2019change, touati2020anomaly, zheng2021unsupervised} or GANs \citep{niu2018conditional} are used as well. 
Self-supervised learning strategies have shown great success recently, including for change detection task. 
In \cite{saha2022self}, the authors take the advantage of the under-representation of changed areas and of the multi-sensor configuration to force the network to learn similar features in patches from the same spatial location and different features for two random patches through a contrastive loss. Contrastive learning is also used at super-pixel level \citep{chen2022self} or to separate features from similar and dissimilar patches generated using an unsupervised image segmentation algorithm \citep{cai2021task}. \cite{leenstra2021self} experimented two different pre-text tasks: overlapping and non-overlapping patches discrimination, and minimizing the difference between overlapping patches in the feature space. Notice that the second task seems to bring better change detection results, this is in line with the work of \cite{saha2022self}. \cite{dong2020self} make use of the discriminator of a GAN trained to differentiate samples from bi-temporal images. When image time series are available, the prediction of the natural order of images seems to be a suitable pre-text task for change detection \citep{saha2020change}. 
Pre-trained models can also be used to generate latent features further transformed in the final change map. Building upon this idea, \cite{saha2019unsupervised} propose to adapt the well-known CVA algorithm \citep{malila1980change} to deep latent features with Deep Change Vector Analysis (DCVA) method. A deep change magnitude coefficient is computed for each pixel from automatically selected deep features. These pixel-wise coefficients, named the  latent change map, are then converted to the final change map through thresholding. Let us also outline that in the literature, different other strategies are experimented to generate the latent change map using features similarity analysis \citep{zhang2016change, chen2022self}, slow features analysis \citep{du2019unsupervised}, features distance combined with mutual information metric \citep{zheng2021unsupervised}, multi-scale feature map fusion \citep{li2022remote}. Thresholding operation is very common to obtain the final change map \citep{liu2016deep, du2019unsupervised, chen2022self,zheng2021unsupervised}, but clustering is also used for binary \citep{zhang2016change,lv2018deep,touati2020anomaly} or multi-class change identification \citep{wu2021unsupervised}. In many use cases, a simple thresholding is enough to achieve interesting results. However, since it is only able to extract binary change information, it cannot deal with more complex scenarios (where various kinds of change are observed or more semantics are needed).

Following the analysis of the state-of-the-art, we propose to rely on DCVA \citep{saha2019unsupervised} to build an unsupervised deep learning method for 3D PCs change detection. Given the interesting results achieved by self-supervised learning for 2D images change detection, we will adapt the idea of \cite{saha2022self} for 3D particular data.


\section{Methodology}
\label{sectionProposedMethod}
\begin{figure}[!t]
\centering
\includegraphics[width=0.5\textwidth]{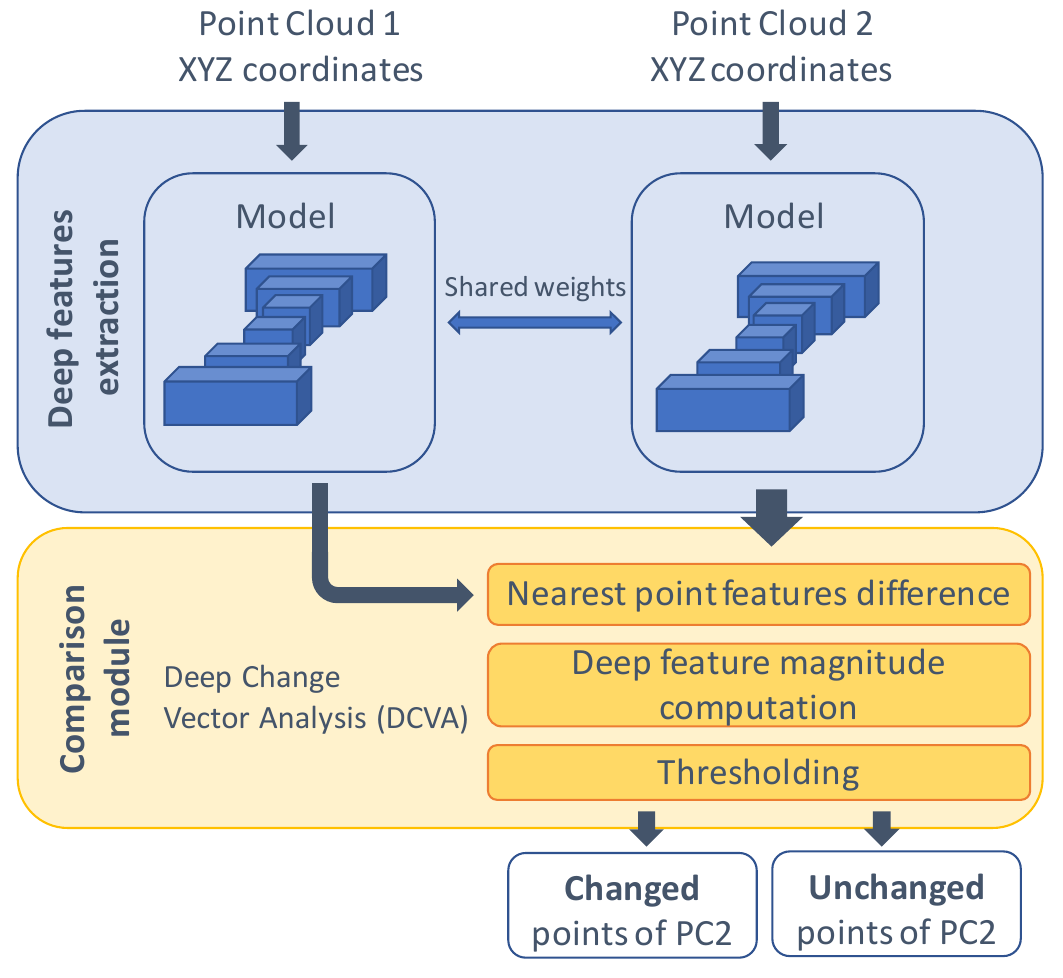}
\caption{Overview of the proposed method}
\label{fig:schemaGal}
\end{figure}

Our proposed method is fully unsupervised and is composed of two major steps, as described in Figure~\ref{fig:schemaGal} and detailed in sections \ref{sec:ssl} and \ref{sec:dcva}, respectively. The first one consists in extracting deep features that will be compared in the second step to extract changes. In the first stage, a network is trained to segment each PC individually using a self-supervised learning strategy.
In this study, to adapt such a  framework to 3D PCs, we use the Kernel Point -- Fully Convolutional Neural Network (KP-FCNN) \citep{thomas2019kpconv}  as the backbone for the deep feature extraction part. Indeed, this network, based on Kernel Point Convolution (KPConv), showed interesting results even when dealing with the remote sensing of large scenes \citep{varney2020dales}. Furthermore, the architecture is similar to 2D architectures, except that 2D convolutions are replaced by KPConv ones. Based on kernel points, these convolutions are specially designed to extract features from 3D PCs.
In the second part, we use DCVA to compare deep features and achieve 3D PC change detection.

 We will denote $\mathcal{P}$ a PC and $\mathcal{F}^{l}$ its associate features at the layer $l \in \left\{ 0 \ldots L\right\}$ of the network symbolized by $f_{\text{KP-FCNN}}$. The index $1$ (resp. $2$) corresponds to the older PC noted $\mathcal{P}_{1}$ (resp. newer PC noted $\mathcal{P}_{2}$) and $N$ denotes the number of points $p$ in the PC $\mathcal{P}$. We assume that $\mathcal{P}_{1}$ and $\mathcal{P}_{2}$ are registered together. To do this, a traditional flowchart like the Iterative Closest Point \citep{besl1992method} algorithm can be used for example.
 
\subsection{Training deep feature extraction: self-supervision}\label{sec:ssl}
Inspired by \cite{saha2022self}, we propose a self-supervised approach that does not require complementary data to train the feature extraction network. While in \cite{saha2022self}, self-supervised learning idea is based on learning to extract similar features from very different SAR and optical acquisitions from a same scene, we thought the variation in 3D points distribution may also be an advantage. Let us note that even in unchanged parts, 3D PCs may have different distributions due to the various acquisition plans, sensors, weather conditions, etc. Although differences in distributions make the direct comparison of PCs impossible, this property can be an asset for training a network to predict similar attributes over an unchanged area regardless of distribution.  

This is the idea of the self-supervised part. The network is trained using three different losses on an unlabeled training set from the same two campaigns of acquisition as the testing set. At each iteration, the back-propagation of the gradient is made using alternatively one of the three losses. 
Thereby, in each iteration, a batch of $\mathcal{B}$ tiles of the older PC, denoted as $\mathcal{X}_{1} =\left\{ x_{1}^{1}, \ldots, x_{1}^{\mathcal{B}} \right\}$, and the corresponding $\mathcal{B}$ tiles of the newer PC, $\mathcal{X}_{2} =\left\{ x_{2}^{1}, \ldots, x_{2}^{\mathcal{B}} \right\}$, are independently given to the network, resulting in features $y$:

\begin{equation}
    y_{1}^{b} = f_{\text{KP-FCNN}}(x_{1}^{b})
\end{equation}
\begin{equation}
    y_{2}^{b} = f_{\text{KP-FCNN}}(x_{2}^{b})
\end{equation}
where $y_{1}^{b}$ and $y_{2}^{b}$ have the dimension $N_{1}^{b} \times K$ and $N_{2}^{b} \times K$, respectively. We recall that $N_{1}^{b}$ and $N_{2}^{b}$ are the number of points in the corresponding tiles. $K$ refers to the dimension of the output which is in practice the number of desired clusters (see the deep clustering loss below). 

\begin{figure*}
    \centering
    \includegraphics[width=0.8\textwidth]{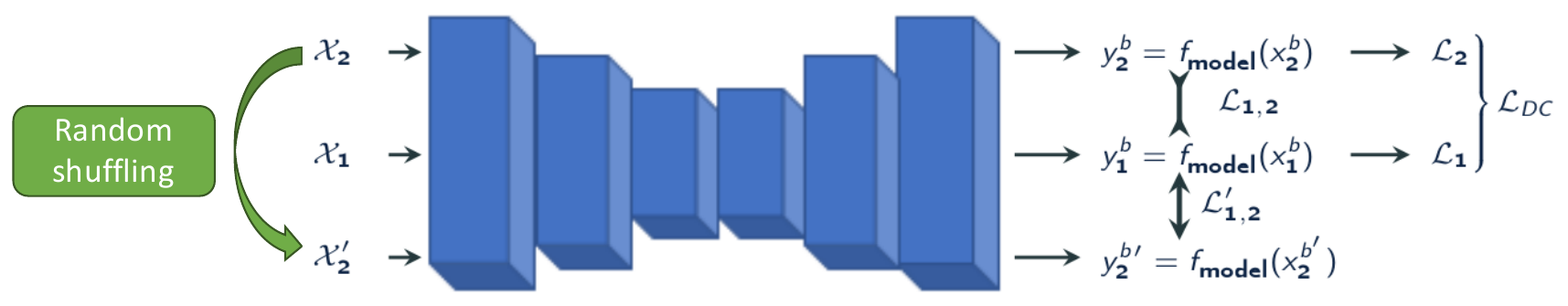}
    \caption{Schema of the self-supervised training of the back-bone. The three different losses are alternatively used to modulate the model weights: the deep clustering loss $\mathcal{L}_{DC}$, the temporal consistency loss $\mathcal{L}_{1,2}$ (with attractive arrows) and the contrastive loss $\mathcal{L}_{1,2}'$ (with repulsive arrows). }
    \label{fig:ssl}
\end{figure*}

As for losses, we alternatively use three different terms, as illustrated in Figure~\ref{fig:ssl}. The first one is based on the deep clustering principle to force the network to learn discriminative features. Deep clustering relies on a pseudo-label assignment which will be used to train the network \citep{caron2018deep}. In this study, pseudo-labels are obtained for each point by taking the argument of the maxima as the output of the network as in \cite{saha2022self}. For example, for each point $i$ of the tile $x_{1}^{b} \in \mathcal{X}_{1}$, the corresponding pseudo-label $c_{1,i}^{b}$ is defined as:

\begin{equation}
    c_{1,i}^{b} = \underset{k \leq K}{\arg \max}~y_{1,i}^{b}(k)
\end{equation}
where $K$ is the number of clusters, which is a hyper-parameter to fix. It can be associated with the number of semantic classes to segment in a single PC (note that this does not concern the number of classes of change between two PCs). However, intuitively, if $K$ is small, learned features will not be discriminative enough as large sets of points will be classified in the same class. On the contrary, with excessively high value, features will be too precise, and no generalization will be possible. In this study, it has been set empirically.

Based on these pseudo-labels, two deep clustering losses $\mathcal{L}_{1}$ and  $\mathcal{L}_{2}$ are defined as the cross-entropy between $(c_{1},y_{1})$ and between $(c_{2},y_{2})$ respectively. The average of these two terms is taken to modulate weights:

\begin{equation}
    \mathcal{L}_{DC} = \frac{\mathcal{L}_{1} + \mathcal{L}_{2}}{2}
\end{equation}
However, with such losses, we observed that the network was collapsing and predicting all the points in a single cluster. 
To prevent this obvious solution, a weighting of the cross-entropy losses was done by applying the following weights:

\begin{equation}
    W_{k} = \frac{1}{\sqrt{\alpha C_{k}}}
\end{equation}
for each cluster $k \in \left\{1, \ldots, K\right\}$, $C_{k}$  being the number of points in the cluster $k$. $\alpha$ is fixed to $K{\sum}_{h=1}^K C_{h}$ as done in the public implementation of KP-FCNN in the Torch-Points3D framework \citep{chaton2020torch}. Weights are recomputed at each epoch. Intuitively, the deep clustering loss enables the network to learn discriminative features to be able to segment each point into clusters. 

In addition to these clustering losses, we add a temporal consistency loss whose rule is to push the network to make similar predictions for tiles from different times but at similar places.  As a matter of fact, even in unchanged areas, the point distribution in 3D point clouds differ from each other. Thus, by assuming that permanent changes between two dates are rare in proportion to the unchanged parts in urban areas, the temporal consistency loss enforces the network to make similar predictions for each point of the newer PC compared to the corresponding nearest point in the older PC. 
Therefore, both predictions ($y_{1}^{b}$ and $y_{2}^{b}$) are ordered before computing the loss:

\begin{equation}
    y_{1_{\text{ordered}}}^{b} = y_{1_{j| j=\arg\min (\| p_{2i} - p_{1j} \|), \forall p_{2i} \in x_{2}^{b}}}^{b} 
\end{equation}
\begin{equation}
    y_{2_{\text{ordered}}}^{b} = y_{2_{i, \forall p_{2i} \in x_{2}^{b}}}^{b} 
\end{equation}
\begin{equation}\label{eq:ltempcons}
    l_{12,i}^{b} = \lVert y_{1_{\text{ordered}},i}^{b} - y_{2_{\text{ordered}},i}^{b} \rVert_{1}
\end{equation}
The temporal consistency loss, $\mathcal{L}_{1,2}$, is then given by taking the mean of $l_{12,i}^{b}$ over all considered points ($p_{2i}^{b} \in \mathcal{X}_{2}$) of all tiles of the batch. Notice that this strategy of nearest point correspondence has already been successfully employed in supervised 3D PC change detection \citep{degelis2023siamese}.

The third loss is a contrastive loss to encourage the network to produce dissimilar features for different tiles. As proposed in \cite{saha2019unsupervised}, the contrastive loss is computed in a similar way to $\mathcal{L}_{1,2}$ by having previously randomly shuffled the batch $\mathcal{X}_{2}$ into $\mathcal{X}_{2}'$ to obtain different tiles between $\mathcal{X}_{1}$ and $\mathcal{X}_{2}'$. The loss $l_{12,i}^{b'}$ is defined as follows:

\begin{equation}
    y_{1_{\text{ordered}}}^{b} = y_{1_{j| j=\arg\min (\| p_{2i}' - p_{1j} \|), \forall p_{2i}' \in x_{2}^{b'}}}^{b} 
\end{equation}
\begin{equation}
    y_{2_{\text{ordered}}}^{b'} = y_{2_{i, \forall p_{2i}' \in x_{2}^{b'}}}^{b'} 
\end{equation}
\begin{equation}
    l_{12,i}^{b'} = - \lVert y_{1_{\text{ordered}},i}^{b} - y_{2_{\text{ordered}},i}^{b'} \rVert_{1}
\end{equation}
Similarly to \cite{saha2019unsupervised}, $\mathcal{L}_{1,2}'$ is given by taking the mean of the exponential of the term $l_{12,i}^{b'}$ over all considered points of all tiles in the batch $\mathcal{X}_{2}'$:
\begin{equation}
    \mathcal{L}_{1,2}' =  \sum_{i \in \mathcal{X}_{2}'} \frac{ e^{l_{12,i}^{b'}}}{N_{\mathcal{X}_{2}'}}
\end{equation}
where $N_{\mathcal{X}_{2}'}$ is the number of points in the batch $\mathcal{X}_{2}'$. Here, the exponential is added to avoid over-penalizing the network when $l_{12,i}^{b'}$ is too far from 0. Indeed, even by shuffling $\mathcal{X}_{2}$, some areas can keep the same semantic, for example there might always be some ground points.

To summarize, the deep clustering loss makes it possible to learn discriminative features, the temporal consistency loss forces the network to predict similar features for similar areas regardless of the point distribution, and the contrastive loss avoids a trivial solution where all predictions are similar for both times. The overall process is given in Algorithm~\ref{alg:ssl} and illustrated in Figure~\ref{fig:ssl}. This method is referred to as Self-Supervised Learning (SSL) in the following part.

\begin{algorithm}
\caption{Self-supervised training of the back-bone}\label{alg:ssl}
\begin{algorithmic}
\STATE Initialize KP-FCNN weights 
\FOR{$e \gets 1$ to $\mathcal{E}$}
\STATE Sample $\mathcal{B}$ tiles from $\mathcal{P}_{1}$, denoted as $\mathcal{X}_{1}$
\STATE Obtain corresponding $\mathcal{B}$ tiles from $\mathcal{P}_{2}$, denoted as $\mathcal{X}_{2}$
\STATE Obtain $\mathcal{X}_{2}'$ as random shuffling of $\mathcal{X}_{2}$
\FOR{$i \gets 0$ to $\mathcal{I}-1$}
\FOR{$b \in \mathcal{B}$}
\STATE $y_{1}^{b} = f_{\text{KP-FCNN}}( x_{1}^{b})$
\STATE $y_{2}^{b} = f_{\text{KP-FCNN}}( x_{2}^{b})$
\STATE $y_{2}^{b'} = f_{\text{KP-FCNN}}( x_{2}^{b'})$
\ENDFOR
\STATE Compute the weights $W_{k}$ considering $y_{1}^{b}$ and $y_{2}^{b}$
\STATE Calculate weighted deep clustering loss $\mathcal{L}_{1}$
\STATE Calculate weighted deep clustering loss  $\mathcal{L}_{2}$ 
\STATE Calculate temporal consistency loss $\mathcal{L}_{1,2}$
\STATE Calculate contrastive loss $\mathcal{L}_{1,2}'$
\IF{$i \bmod{3} = 0$}
\STATE Use $\mathcal{L}_{DC} = \frac{\mathcal{L}_{1} + \mathcal{L}_{2}}{2}$ to modulate KP-FCNN weights
\ELSIF{$i \bmod{3} = 1$}
\STATE Use $\mathcal{L}_{1,2}$ to modulate KP-FCNN weights
\ELSE 
\STATE Use $\mathcal{L}_{1,2}'$ to modulate KP-FCNN weights
\ENDIF
\ENDFOR
\ENDFOR
\end{algorithmic}
\end{algorithm}

\subsection{Deep feature comparison}\label{sec:dcva}
Once a model is trained to perform a segmentation task, it can be used on both input PCs to extract features at different levels of abstraction and complexity depending on the layer. These extracted features can be used in order to highlight changes applying the Deep Change Vector Analysis (DCVA) principle, initially developed for 2D pixel change retrieving through deep features comparison \citep{saha2019unsupervised}. As shown in the comparison module of Figure~\ref{fig:schemaGal}, the point-wise change identification is realized by taking the magnitude of the difference ($\delta^{l}$) between feature vectors computed for each point of the newer PC, $p_{2i} \in \mathcal{P}_{2}$, with the nearest point of the older PC, $p_{1j} \in \mathcal{P}_{1}$. In other words, the feature difference $\delta^{l}$ is computed between features $\mathcal{F}_{1}^{l}$ and $\mathcal{F}_{2}^{l}$ for each PC, $\mathcal{P}_{1}$ and $\mathcal{P}_{2}$ respectively, according to the following equation:

\begin{equation}\label{eq:npdiff}
    \delta^{l}_{i}= f_{2i}^{l} - f^{l}_{1j | j=\arg\min (\| p_{2i} - p_{1j} \|)}
\end{equation}
with $f_{1j}^{l} \in \mathcal{F}_{1}^{l}$, $f_{2i}^{l} \in \mathcal{F}_{2}^{l}$, $p_{1j} \in \mathcal{P}_{1}$ and $p_{2i} \in \mathcal{P}_{2}$.
The magnitude of the difference, also called deep feature magnitude coefficient, is obtained by taking the L2-norm of $\delta^{l}$.
A threshold is applied on the deep feature magnitude coefficient to distinguish between changed and unchanged points. As in \cite{saha2019unsupervised}, the threshold is selected using the unsupervised Otsu algorithm \citep{otsu1979threshold}. Let us remark that Otsu thresholding is commonly used for change detection, for both 2D \citep{saha2019unsupervised, du2019unsupervised, zheng2021unsupervised} and 3D \citep{dos2021use, marmol2023analysis}. The trained network extracts similar features for two similar areas, thus the deep magnitude coefficient is close to zero in the unchanged part. 
The choice of the layer from which features are taken is a hyper-parameter to be set.  Combined with SSL, the method is called SSL-DCVA.

\section{Experiments and discussion}\label{sectionExperiments}
\subsection{Dataset}
We performed the same experiments on a real dataset consisting of two different ALS campaigns throughout the Netherlands. We extracted some tiles for the training and the test of the method among the publicly available dataset Actueel Hoogtebestand Nederland (AHN). This dataset consists of a total of four surveys throughout the entire country \citep{sande2010assessment}, allowing multi-date change extraction \citep{CSEREP2023103174}. The two last surveys (AHN3 and AHN4) have been semi-automatically annotated to assign a semantic label to each point. Five different classes are distinguished:
ground, buildings, water, civil engineering structures (e.g., bridges) and clutter. This mono-date segmentation allowed us to manually derive a change ground truth for the test set. Concerning the changes, we decided to include in change areas the new or demolished building (in this case the building footprint on the ground points is marked as changed), new vegetation and new clutter. 
Since our focus is on 3D object changes, our manual annotation was ignoring changes in land cover if not characterized by modifications of the geometry. As such, changes like grassland to bare soil are only discernible using red-green-blue (RGB) color information, and do not imply 3D geometric changes. Besides, let us notice that the RGB data provided in AHN have not been acquired at the same date as LiDAR PCs implying numerous disagreement between 3D geometry and RGB information of changed scenes. Therefore, we explicitly discard RGB information.

As far as training PCs are concerned, they are sub-parts of tiles 31HN1\_22, 31HN1\_23, 31HZ1\_04, from the divided AHN dataset (\url{https://geotiles.nl/}). Concerning the test set, qualitative analysis was performed on a sub-part of tile 37EN1\_08 (see Figure~\ref{fig:resG}(a,c)).  This sub-part is selected to show sufficient change instances to properly evaluate the method. AHN data does not contain any change annotation. Thus, deriving change labels from available semantic labels is not obvious due to point cloud characteristics (e.g., no point to point corresponding). Thereby, we selected a surface of 12,400 m$^{2}$ from the tile 37EN1\_08 and conducted a manual annotation. Even if here the change detection is only binary, some different types of changes are present in this area including new buildings, demolished buildings, new vegetation and clutter. 

The density of AHN3 is about 12 points/m$^2$ while AHN4 is about 22 points/m$^2$. Height and planimetric stochastic errors are 5 cm. In addition to point coordinates, AHN data also includes LiDAR intensity and the number of returns. However, since here we focus on raw 3D PCs only, we rely solely on the 3D point coordinates to feed the network.

\subsection{Experimental protocol}
To assess our proposed self-supervised strategy to extract relevant deep features, we will compare our results with those obtained by our 3D PC DCVA adaptation using a network pre-trained on an annex task such as semantic segmentation using labels from a publicly available dataset as done in the 2D case in \citep{saha2019unsupervised}.
Indeed, while datasets annotated according to the change are not common when dealing with 3D PCs, public datasets with a mono-date semantic annotation in urban environment are widely spread \citep{hackel2017semantic3d,roynard2017parislille3d,varney2020dales,kolle2021hessigheim}. By choosing a public dataset as close as possible to the unlabeled change detection dataset to perform supervised training of the network, one could expect that extracted features will be consistent in unchanged areas and different in the case of changes. 
In our study, we use the Hessigheim 3D (H3D) Aerial LiDAR Survey to train the network for semantic segmentation. The H3D dataset consists of four different PCs at various dates and comes with labels related to 11 semantic classes that have been manually annotated \citep{kolle2021hessigheim}. In practice, the training is performed on H3D PCs acquired in March 2016 on behalf of the national mapping agency of Baden-Württemberg, Germany. This survey has a mean point density of about 20 points/m$^2$. In the following study, we refer to this method by Supervised Semantic Segmentation Training (SSST). Combined with DCVA, the method is called SSST-DCVA.

Additionally, even if no specific unsupervised deep method exists so far in the literature for 3D change detection, we decided to compare with the supervised Siamese KPConv network \citep{degelis2023siamese} trained on the simulated Urb3DCD dataset \citep{degelis2021change} and directly applied to AHN-CD testing set without any retraining. Thereby, this strategy referred as Siamese KPConv transfer is another unsupervised baseline as no label from the target dataset (i.e., AHN-CD) is used during the training. We also provide results achieved with a pure supervised baseline, training Siamese KPConv on the AHN-CD train set (with automatic annotation of change).
To further benchmark our method against the state-of-the-art, we provide a comparison with Cloud-to-Cloud (C2C) \citep{girardeau2005change} and multi-scale model-to-model cloud comparison (M3C2) \citep{lague2013accurate} distance-based methods. These methods constitute unsupervised baselines for 3D point-based binary change detection \citep{shirowzhan2019comparative}. In particular, to obtain final binary change information, a thresholding based on Otsu algorithm \citep{otsu1979threshold} is applied on the Hausdorff point-to-point distance computed in C2C.
The traditional M3C2 method uses local surface normal and orientation to compute the 3D distance between two PCs \citep{lague2013accurate}. This method relies on statistical tests on distances between local surface normal and orientation features of the two PCs to automatically extract significant changes. In addition, we provide a comparison with an unsupervised machine learning method, namely a $k$-means, trained on 3D hand-crafted features proposed by \cite{tran2018integrated}.

\subsection{Experimental settings}

For the same computational reason as 2D images are divided into patches, original PCs are also divided into tiles to run deep learning experiments. As done in the supervised deep framework for 3D PC change detection from \cite{degelis2023siamese}, vertical 3D cylinders are chosen to make sure that whatever changes occur between the two dates, at least the ground is visible. Also, considering that the chosen back-bone model is KP-FCNN, which is relying on Kernel Point Convolution, a first sub-sampling rate has to be chosen \citep{thomas2019kpconv}. Let us remark that the choice of the radius of cylinders and of the first sub-sampling rate has to be made in consideration of the scale of the sought changes. The thinner the first sub-sampling rate is, the more points from the original PC will be given to the network to extract useful features. The larger the radius is, the more context will be taken into account. Thereby a compromise between a large radius size and a thin sub-sampling rate has to be done due to memory limitations. We set the parameters following previous studies \citep{degelis2023siamese} and use the same first sub-sampling rate value. As far as the radius is concerned, we had to change the value for the AHN-CD dataset to fit the memory constraints. 

Concerning SSL, the training is realized using cylinders of 20~m in radius and a first sub-sampling rate of 0.5~m.
Then, the best results are obtained using 6 clusters for the deep clustering loss, and after 15 epochs of training. One hundred cylinders are used for each epoch, with a batch size of 10.

For the SSST part, KP-FCNN is trained using cylinders of 10~m in radius with a first sub-sampling rate of 0.2~m. A total of 6,000 cylinders are used for the training at each epoch. The batch size is 10. The 11 H3D classes are fused into 7 classes better suited for the target datasets. It requires around 90 epochs to converge.

Networks are optimized using a Stochastic Gradient Descent with a momentum of 0.98. The learning rate is set at 0.01 and decreases exponentially. These settings were chosen on the basis of the initial publication of KP-FCNN back-bone \citep{thomas2019kpconv}. Regardless of the first sub-sampling rate, the final results are given at the original resolution. Indeed, a reprojection step is performed with regard to the nearest neighbor in order to obtain a final result for each point of the original PC.

For the choice of the layer to take features for the DCVA, several configurations have been tested. Knowing that KP-FCNN has 9 layers, the best results, reported here, are obtained using the 7th layer for SSST and the 8th for the SSL strategy.

Following the thresholding step, a cleaning of isolated predictions is realized to spatially smooth the results. Note that this cleaning step is systematically applied to ensure fair comparison.

Experiments were conducted with a single Graphic Processing Unit (GPU) (Nvidia Tesla V100 SXM2 16 GB).

\subsection{Results and discussion}
\begin{figure*}[htbp]
   \centering
 \begin{minipage}[t]{0.29\linewidth}
  \centering
  \centerline{\includegraphics[trim=0.8cm 0.3cm 0.cm 0.9cm, clip, width=\linewidth]{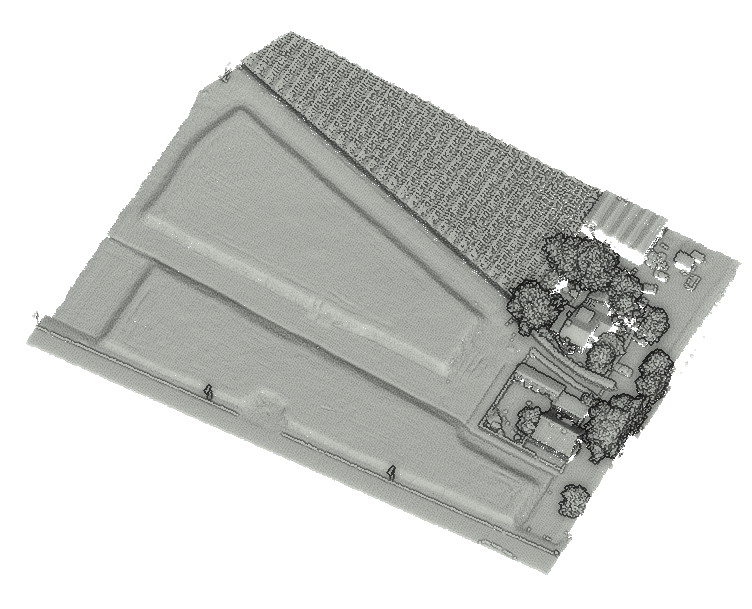}}
  \centerline{ }
  \centerline{a)  AHN3 data (time 1)}\medskip
 \end{minipage}
  \begin{minipage}[t]{0.29\linewidth}
  \centering
  \centerline{\includegraphics[trim=0.8cm 0.3cm 0.cm 0.9cm, clip, width=\linewidth]{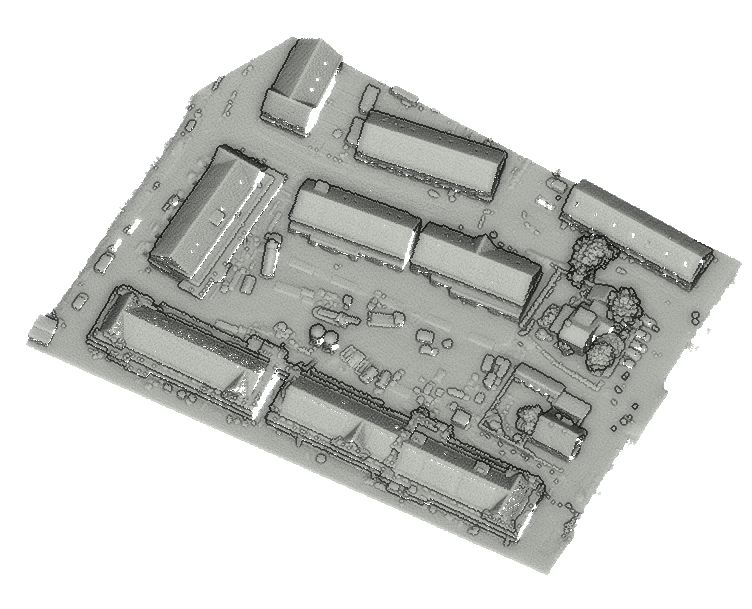}}
  \centerline{ }
  \centerline{b)  AHN4 data (time 2)}\medskip
   \end{minipage}
\begin{minipage}[t]{0.29\linewidth}
  \centering
  \centerline{\includegraphics[trim=0.8cm 0.3cm 0.cm 0.9cm, clip, width=\linewidth]{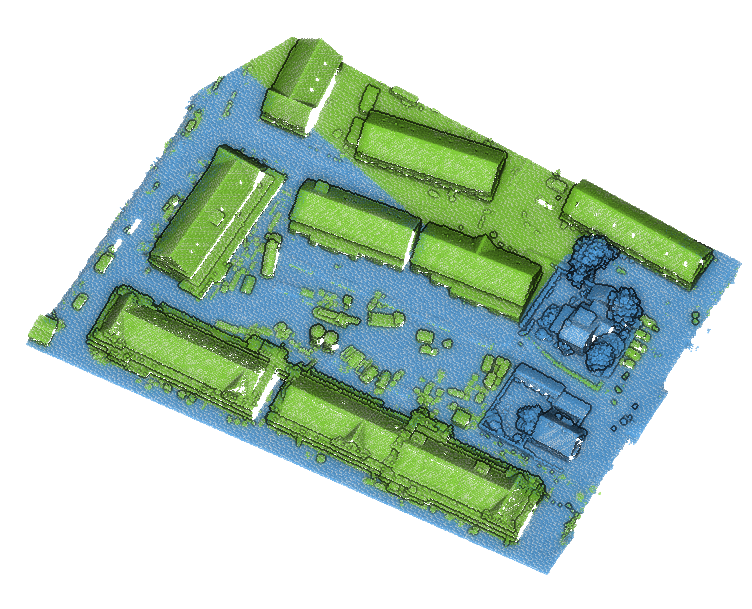}}
  \vspace{-0.2cm}
  \begin{tikzpicture}
    		\begin{axis}[
            		xmin=1,
                    xmax=2,
                    ymin=1,
                    ymax=2,
                     hide axis,
    				width=0.5\linewidth ,
    				mark=circle,
    				scatter,
    				only marks,
    				legend entries={\textcolor{black}{Unchanged}, \textcolor{black}{Changed}},
    				legend cell align={left},
    				legend style={draw=none,at={(0,0),fill=none}, legend columns=2,/tikz/every even column/.append style={column sep=0.2cm}}]
    			\addplot[TN] coordinates {(0,0)}; 
    			\addplot[TP] coordinates {(0,0)};
    		\end{axis}
	    \end{tikzpicture} 
     \vspace{-0.5cm}
     \centerline{c)  Ground Truth}\medskip
 \end{minipage}
 
 \begin{minipage}[t]{0.32\linewidth}
  \centering
  \centerline{\includegraphics[trim=0.8cm 0.8cm 0.cm 0.9cm, clip,width=\linewidth]{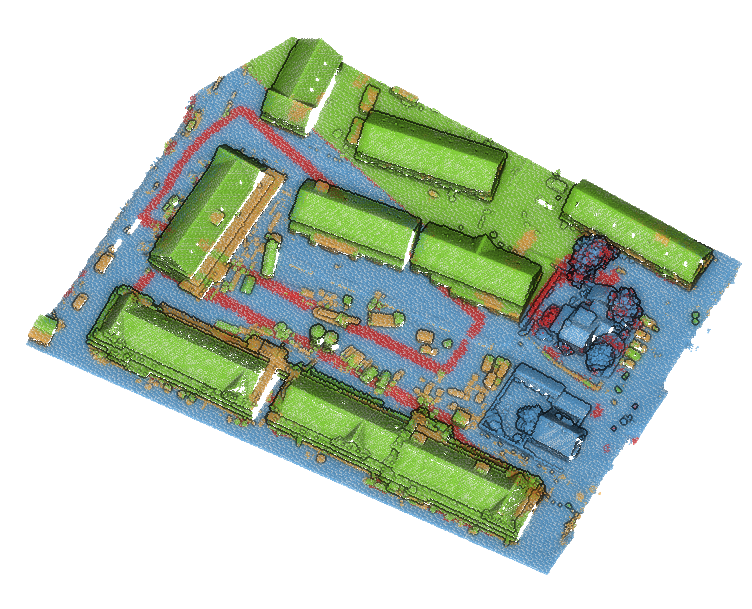}}
  \centerline{d)  SSL-DCVA}\medskip
 \end{minipage}
  \begin{minipage}[t]{0.32\linewidth}
  \centering
  \centerline{\includegraphics[trim=0.8cm 0.8cm 0.cm 0.9cm, clip, width=\linewidth]{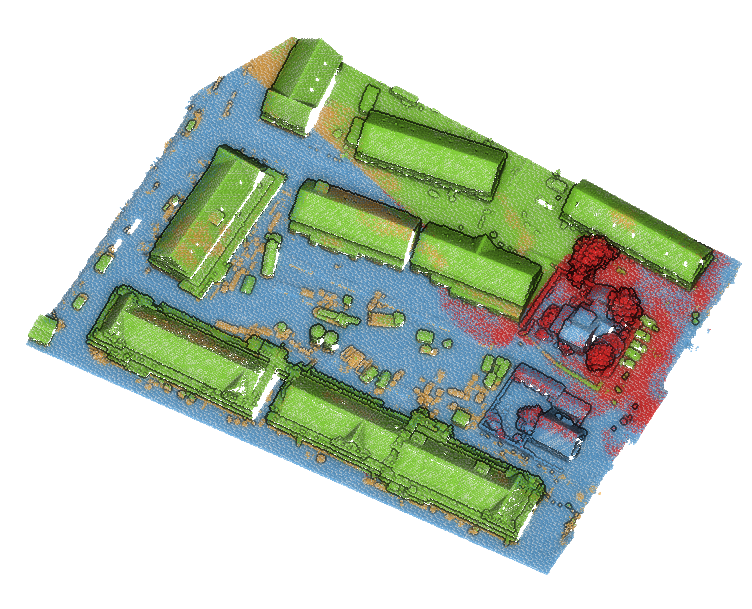}}
  \centerline{e)  SSST-DCVA}
  \centerline{}\medskip
 \end{minipage}
 
\begin{minipage}[t]{0.32\linewidth}
  \centering
  \centerline{\includegraphics[trim=0.8cm 0.8cm 0.cm 0.9cm, clip, width=\linewidth]{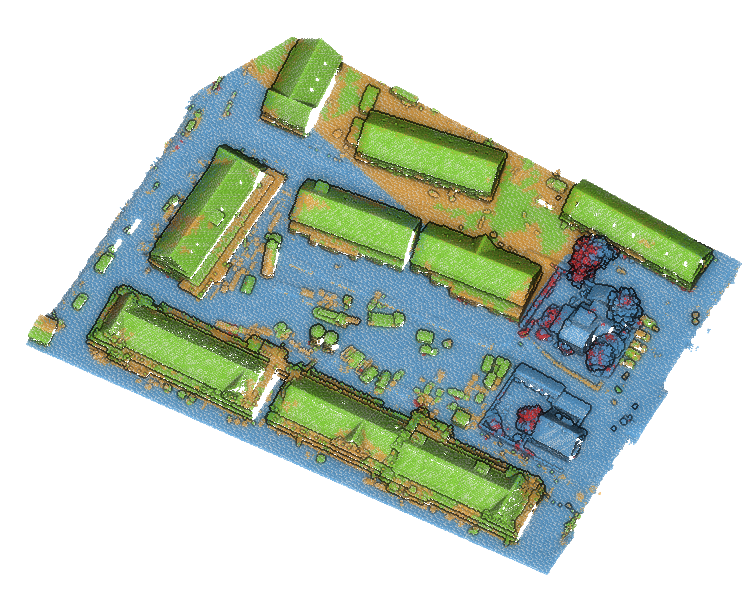}}
  \centerline{f)  Siamese KPConv transfer }
  \centerline{\citep{degelis2023siamese}}\medskip
 \end{minipage}
 \begin{minipage}[t]{0.32\linewidth}
  \centering
  \centerline{\includegraphics[trim=0.8cm 0.8cm 0.cm 0.9cm, clip, width=\linewidth]{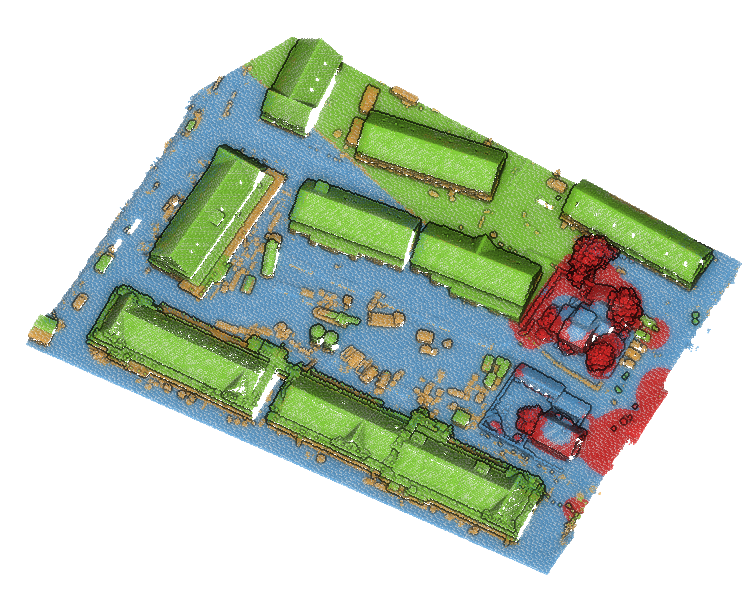}}
  \centerline{g)  $k$-means with hand-crafted features }
  \centerline{from \cite{tran2018integrated}}\medskip
 \end{minipage}

  \begin{minipage}[t]{0.32\linewidth}
  \centering
  \centerline{\includegraphics[trim=0.8cm 0.8cm 0.cm 0.9cm, clip, width=\linewidth]{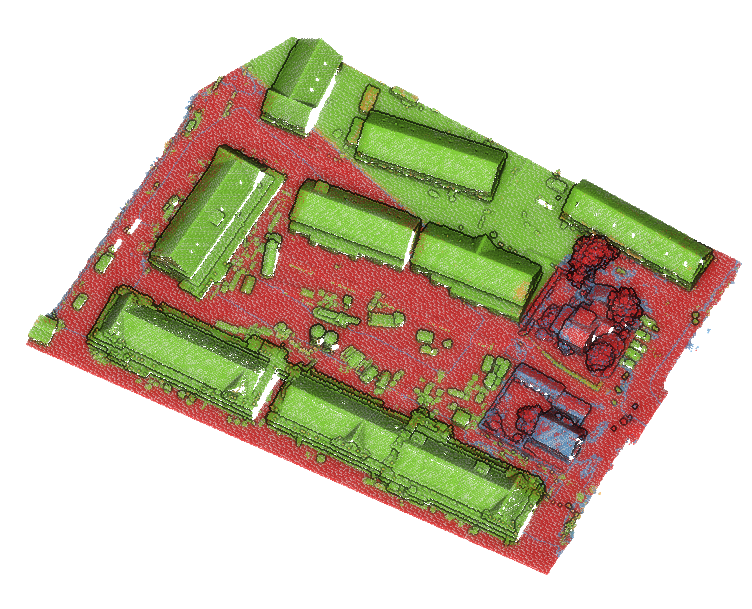}}
  \centerline{h)  M3C2  }
  \centerline{\citep{lague2013accurate}}\medskip
 \end{minipage}
 \begin{minipage}[t]{0.32\linewidth}
  \centering
  \centerline{
   \begin{tikzpicture}
    \node (image) at (0,0) {\includegraphics[trim=0.8cm 0.8cm 0.cm 0.9cm, clip,width=\linewidth]{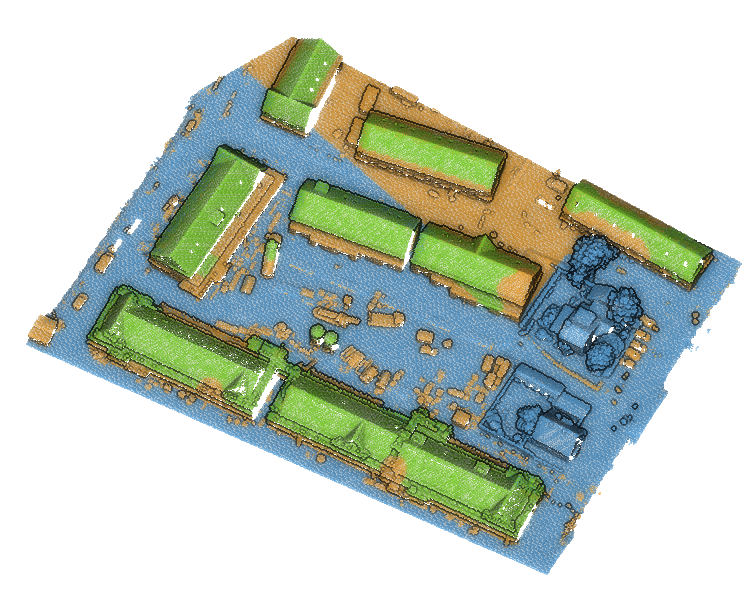}};
       \draw[black, very thick,rotate=50] (1.,-0.8) ellipse (19pt and 14pt);
      \draw[black, very  thick,rotate=95] (-1.2,0.1) ellipse (6pt and 6pt);
      \draw[black, very  thick,rotate=95] (-0.5,1.4) ellipse (6pt and 6pt);
 \end{tikzpicture}}
  \centerline{i)  C2C}
  \centerline{ \citep{girardeau2005change}}\medskip
 \end{minipage}
  \begin{minipage}[t]{0.32\linewidth}
  \centering
  \centerline{  
    \includegraphics[trim=0.8cm 0.8cm 0.cm 0.9cm, clip,width=\linewidth]{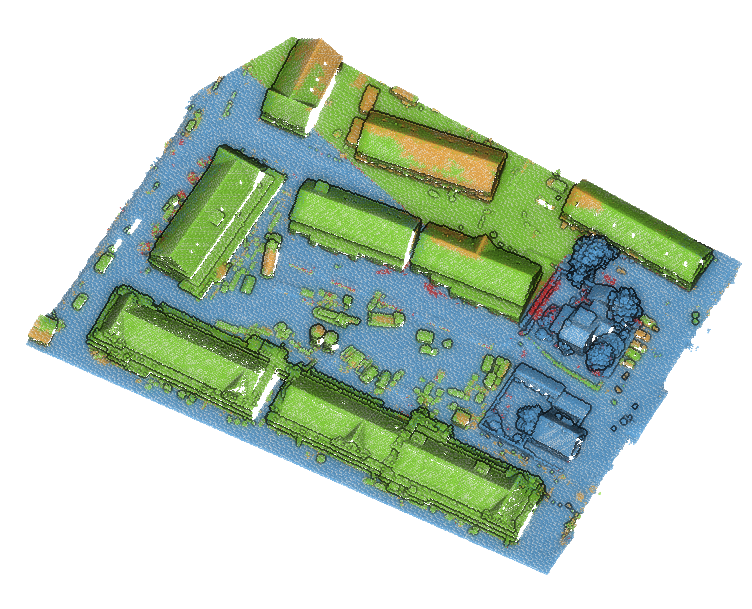}}
  \centerline{j)  Siamese KPConv (supervised)}
  \centerline{ \citep{degelis2023siamese}}\medskip
 \end{minipage}
 
 \begin{minipage}[t]{0.8\linewidth}
  \centering
  \begin{tikzpicture}
    		\begin{axis}[
            		xmin=1,
                    xmax=2,
                    ymin=1,
                    ymax=2,
                     hide axis,
    				width=0.5\linewidth ,
    				mark=circle,
    				scatter,
    				only marks,
    				legend entries={\textcolor{black}{True Negative}, \textcolor{black}{True Positive}, \textcolor{black}{False Negative}, \textcolor{black}{False Positive}},
    				legend cell align={left},
    				legend style={draw=lightgray,at={(0,0)}, legend columns=4,/tikz/every even column/.append style={column sep=0.2cm}}]
    			\addplot[TN] coordinates {(0,0)}; 
    			\addplot[TP] coordinates {(0,0)};
                \addplot[FN] coordinates {(0,0)}; 
    			\addplot[FP] coordinates {(0,0)};
    		\end{axis}
	    \end{tikzpicture} 
 \end{minipage} 
 
    \caption{Qualitative results on the manually annotated testing set. 
    Even if not perfect, learning based methods (SSL-DCVA (d), SSST-DCVA (e), Siamese KPConv transfer (f), $k$-means (g)) seem to better retrieve changes in small objects than distance based methods (M3C2 (h) and C2C (i)). Fully supervised results obtained with Siamese KPConv network are given in (j) for comparison purpose.}
    \label{fig:res}
\end{figure*}
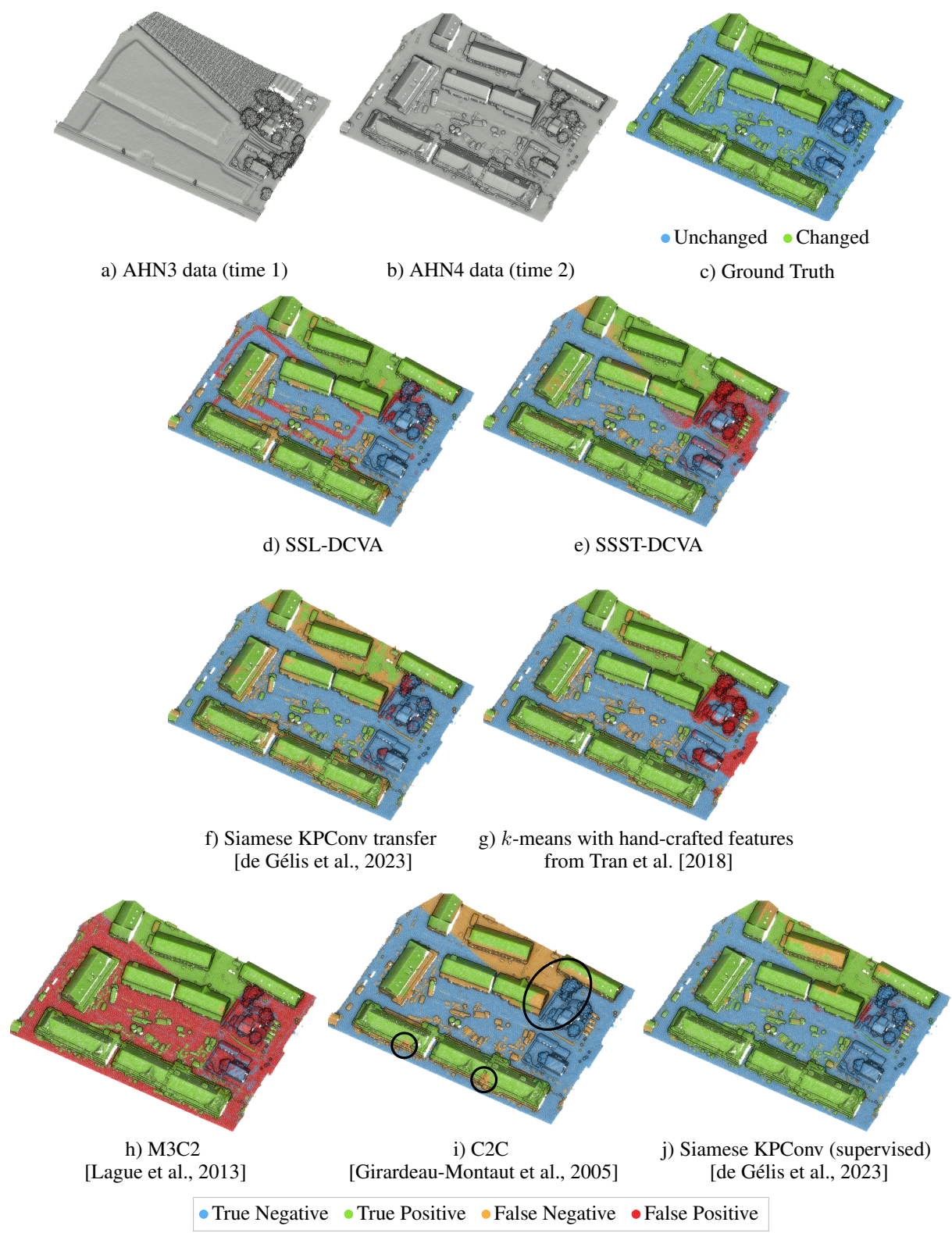

\begin{figure*}[htbp]
   \centering
\begin{minipage}[t]{0.245\linewidth}
  \centering
  \centerline{\includegraphics[width=\linewidth]{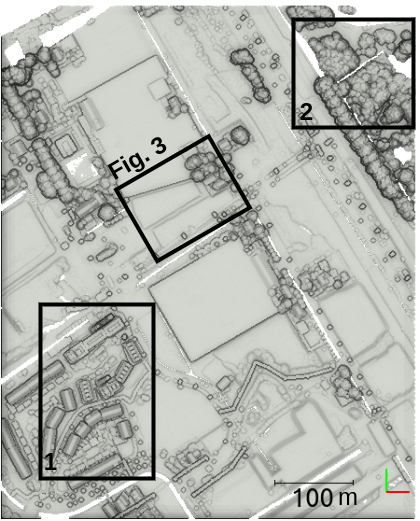}}
   \centerline{a) AHN3 data (time 1)}\medskip
 \end{minipage}
 \begin{minipage}[t]{0.245\linewidth}
  \centering
  \centerline{\includegraphics[width=\linewidth]{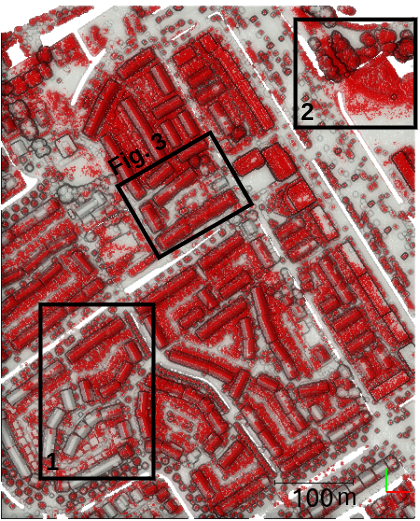}}
  \centerline{b) SSST-DCVA results}\medskip
 \end{minipage}
 \begin{minipage}[t]{0.245\linewidth}
  \centering
  \centerline{\includegraphics[width=\linewidth]{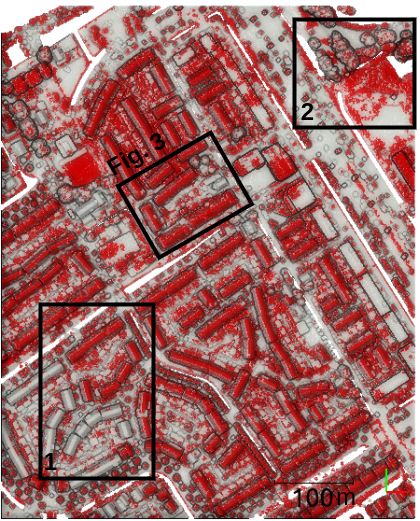}}
  \centerline{c) SSL-DCVA results}\medskip
 \end{minipage}
 \begin{minipage}[t]{0.245\linewidth}
  \centering
  \centerline{\includegraphics[width=\linewidth]{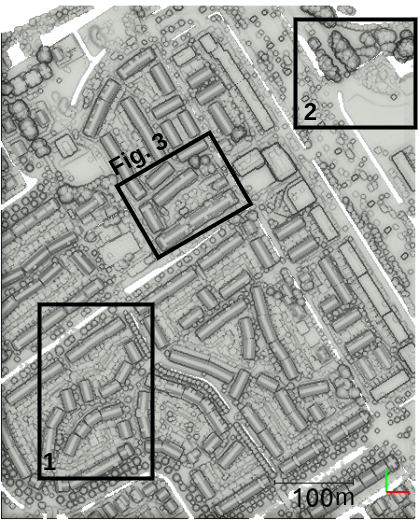}}
  \centerline{d) AHN4 data (time 2)}\medskip
 \end{minipage}
 \\

 \begin{minipage}[t]{0.245\linewidth}
  \centering
  \centerline{\includegraphics[width=\linewidth]{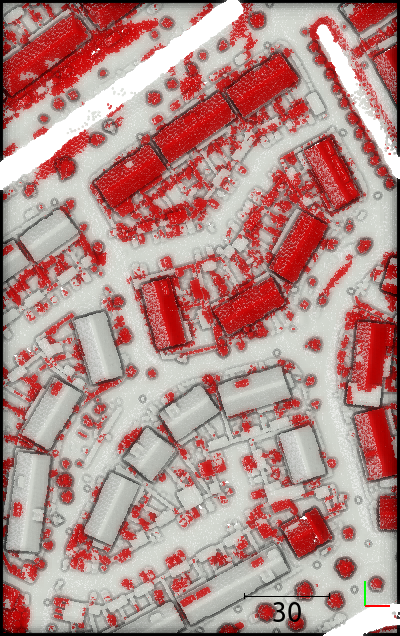}}
  \centerline{e) Zoom 1: SSL-DCVA}
  \centerline{}\medskip
 \end{minipage}
 \begin{minipage}[t]{0.245\linewidth}
  \centering
  \centerline{\includegraphics[trim=0cm 0cm 0cm 0cm, clip, width=\linewidth]{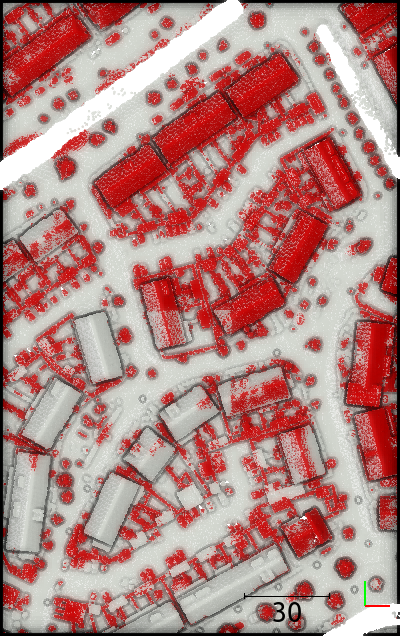}}
  \centerline{f) Zoom 1: SSST-DCVA}
  \centerline{}\medskip
 \end{minipage}
  \begin{minipage}[t]{0.245\linewidth}
  \centering
  \centerline{\includegraphics[trim=0cm 0cm 0cm 0cm, clip, width=\linewidth]{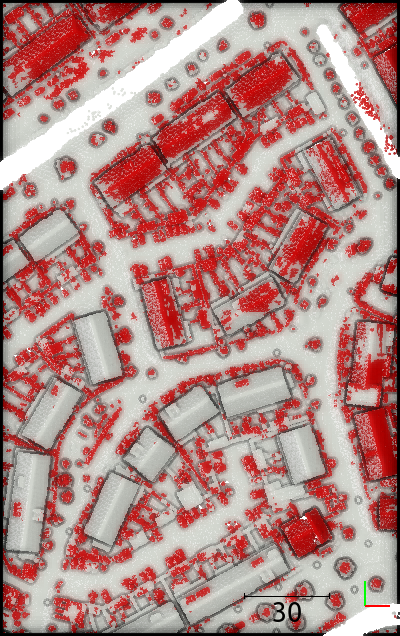}}
  \centerline{g) Zoom 1: Siamese KPConv }
  \centerline{transfer \footnotesize{\citep{degelis2023siamese}}}\medskip
 \end{minipage}
   \begin{minipage}[t]{0.245\linewidth}
  \centering
  \centerline{\includegraphics[trim=0cm 0cm 0cm 0cm, clip, width=\linewidth]{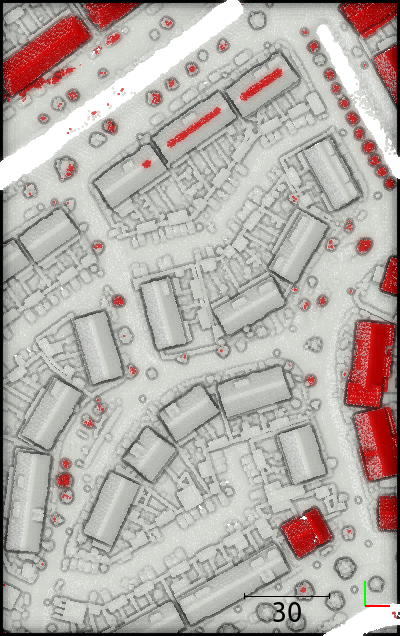}}
  \centerline{h) Zoom 1: C2C}
\centerline{\footnotesize{\citep{girardeau2005change}}}\medskip
 \end{minipage}
\\
 \begin{minipage}[t]{0.245\linewidth}
  \centering
  \centerline{\includegraphics[trim=0cm 0cm 0cm 0cm, clip,width=\linewidth]{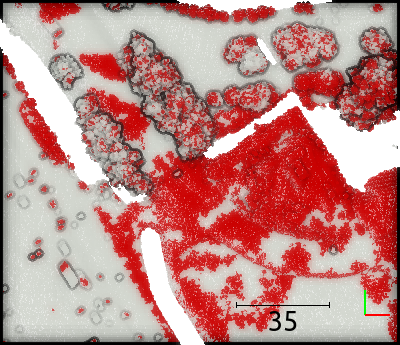}}
  \centerline{i) Zoom 2:  SSL-DCVA}
  \centerline{}\medskip
 \end{minipage}
 \begin{minipage}[t]{0.245\linewidth}
  \centering
  \centerline{\includegraphics[width=\linewidth]{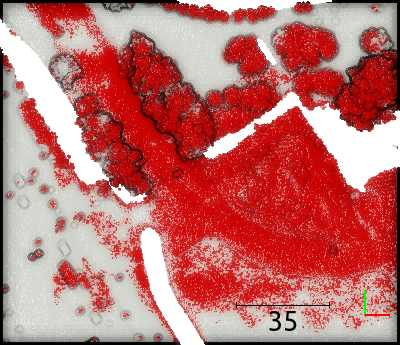}}
  \centerline{j) Zoom 2: SSST-DCVA}
  \centerline{}\medskip
 \end{minipage}
  \begin{minipage}[t]{0.245\linewidth}
  \centering
  \centerline{\includegraphics[width=\linewidth]{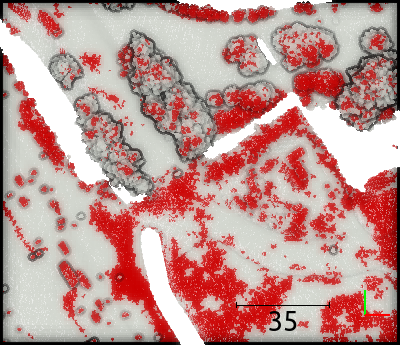}}
  \centerline{k) Zoom 2: Siamese KPConv}
  \centerline{transfer \footnotesize{\citep{degelis2023siamese}}}\medskip
 \end{minipage}
  \begin{minipage}[t]{0.245\linewidth}
  \centering
  \centerline{\includegraphics[width=\linewidth]{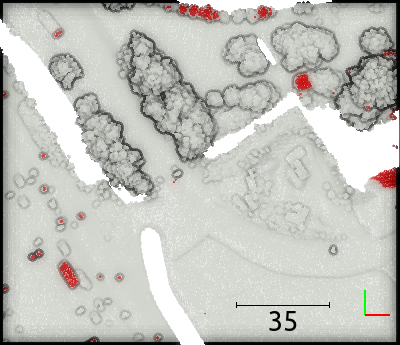}}
  \centerline{l) Zoom 2: C2C}
  \centerline{\footnotesize{\citep{girardeau2005change}}}\medskip
 \end{minipage}
\\
  \begin{tikzpicture}
    		\begin{axis}[
            		xmin=1,
                    xmax=2,
                    ymin=1,
                    ymax=2,
                     hide axis,
    				width=0.5\linewidth ,
    				mark=circle,
    				scatter,
    				only marks,
    				legend entries={\textcolor{black}{Unchanged}, \textcolor{black}{Changed}},
    				legend cell align={left},
    				legend style={draw=lightgray,at={(0,0)}, legend columns=2,/tikz/every even column/.append style={column sep=0.2cm}}]
    			\addplot[unchanged] coordinates {(0,0)}; 
    			\addplot[changed] coordinates {(0,0)};
    		\end{axis}
	    \end{tikzpicture} 
    \caption{Qualitative results on the testing set. Changes are indicated in red. For more precise results, zoom 1 and 2 are visible in e) to l), while zoom Fig. 3 corresponds to the manually annotated area presented in Figure~\ref{fig:res}.}
    \label{fig:resG}
\end{figure*}

Quantitative results are given in Table~\ref{tab:res}. The mean of accuracy (mAcc), the mean of intersection over union (IoU) and the IoU for both changed and unchanged classes are given. Corresponding qualitative results on the manually annotated test set are presented in Figure~\ref{fig:res}. To complete the qualitative analysis of the results, a larger scene has been visually inspected to understand the behavior of our methods in multiple conditions (see Figure~\ref{fig:resG}).

As can be seen in Table~\ref{tab:res}, SSL-DCVA outperforms other methods, including SSST-DCVA. It is worth noting that despite its simplicity (no training is required), C2C provides relevant results. However, the point-to-point distance seems limited in places where for example trees have been replaced by a building of approximately the same height (see regions of interest in Figure~\ref{fig:res}(i)) or where a new building replaced an old one as in the top of zoom 1 where buildings in AHN3 and AHN4 are very different (Figure~\ref{fig:resG}(a,d,h)). Conversely to C2C, M3C2 provides inconsistent results here (see Figure~\ref{fig:res}(h)): the ground elevation has changed slightly between the two acquisitions, so almost all areas are marked as changed. Notice that even when removing ground points for metric computation, the M3C2 method is lagging behind other methods (still about 10\% of mAcc behind the SSST-DCVA algorithm evaluated under the same conditions). Moreover, in this study, we aim at detecting changes in object semantics (new buildings, demolition, new vegetation, etc.), so a change in the ground height is not of interest to us and has not been marked as changed in the ground truth. To further explain the relative poor results of M3C2, we recall that this method was originally developed in a geoscience context to detect changes at different scales, including centimetric \citep{lague2013accurate}. Thereby, it might be inadequate for urban environments, as already noticed in \cite{degelis2021change}.  A distance-based method may not distinguish between topographic and semantic changes as long as the geometry of objects has changed.  Learning-based methods (see Figure~\ref{fig:res}(d-g)) seem to also retrieve  small changed objects, which is not possible with distance-based methods without including too many changes. 

\begin{table*}[!t]
\centering
\begin{tabular}{cc|cc|cc|cc}
& &mAcc &  mIoU&\multicolumn{2}{c|}{IoU (\%)}& \multicolumn{2}{c}{Computation time}\\
& & (\%) &  (\%) & Unchanged & Changed & Training  &  Testing\\
\hline
\multirow{6}{*}{\rotatebox{90}{Unsupervised}}
&SSL-DCVA (ours) & \textbf{85.20} & \textbf{74.14} &\textbf{78.91} & \textbf{69.38} & 9 min & 40 sec\\
&SSST-DCVA (ours) & 81.88 & 66.93 &70.02 &63.85 & 17 hours & 40 sec\\
&Siamese KPConv transfer \citep{degelis2023siamese}& 81.83& 69.76 & 75.80 & 63.73 & 28 hours & 25 sec\\
 & $k$-means (features from \cite{tran2018integrated}) & 81.00 & 66.81 & 71.11 & 62.51 & 40 min &  3 min \\
&M3C2 \citep{lague2013accurate}& 51.77 & 43.56 & 3.66 & 39.90 & - & 5 sec\\

&C2C \citep{girardeau2005change} & 76.67 & 65.16 &76.98 & 53.34 & - & 5 sec \\

\hline
\multirow{1}{*}{\rotatebox{0}{Sup.}} 
&Siamese KPConv  \citep{degelis2023siamese}& 94.23 & 89.96 & 92.27 & 87.65& 15 hours & 25 sec \\
\end{tabular}
\caption{Quantitative results on AHN-CD dataset with both unsupervised and supervised (sup.) methods. Approximate computation times are provided for the training and testing (on the manually annotated part) step.}\label{tab:res}
\end{table*}

By looking at the results of the Siamese KPConv change detection network with transfer onto AHN dataset from Urb3DCD simulated dataset, we can see that performances are quite similar to the SSST-DCVA but are overtaken by SSL-DCVA method. In particular, some differences with the ground truth are visible in the demolished area and at object boundaries. This is probably due to the difference of building types present in the selected area of AHN dataset. Indeed, Urb3DCD dataset contains buildings from a french city center different from train and test areas, as for example, AHN3 data (time 1) contains a glasshouse. The same problem occurs with SSST-DCVA methods, since H3D PCs have different resolution and quality than AHN PCs. This shows the advantage of training directly on a dataset with similar properties to the test set and using recent developments in self-supervised learning. However, when compared to the supervised Siamese KPConv network, unsupervised methods can still largely be improved. The main differences of SSL-DCVA with the ground truth are visible on small objects such as vehicles, road signs or vegetation (see Figure~\ref{fig:res}(c) and (d)). Furthermore, as can be seen in the buildings on the left side of Figure~\ref{fig:res} and right side of Figure~\ref{fig:resG}(c), some omissions remain on new buildings with a flat roof. When looking at the mono-date segmentation of the PC realized before the DCVA step, one can see that flat roofs are classified in the same class as ground so, when comparing features, no changes are highlighted. This raises the difficulty of late-fusion change identification. Indeed, errors in the feature extraction step are propagated in the comparison step. Finally, some false detections are visible on the ground, forming a large trapezium (see Figure~\ref{fig:res}(d)). This is due to changes in the orientation of the ground surface. 

Our method encounters difficulties in unchanged vegetated areas (see the top of zoom 2 in Figure~\ref{fig:resG}(i)) certainly because of the complexity of LiDAR data in such areas with a high variation of point distribution even without changes in the semantics of objects. This results in a mixture of points predicted as changed and unchanged. Furthermore, these vegetated areas may have grown, and the acquisition not realized in the same season implies some differences on the 3D representation of trees. Note that the same problems occur with the other learning-based methods (see the top of zoom 2 in Figure~\ref{fig:resG}(j,k). Looking at Figure~\ref{fig:resG}(l), we can observe that C2C method is not better in this zoom where the vegetation has been removed. Indeed, in AHN3 some points are acquired from the ground to the top of the tree canopy thanks to the LiDAR sensor, thereby the point-to-point distance is not an efficient indicator for changes. Finally, it seems that SSL-DCVA (as well as SSST-DCVA) is more prone to commission than omission changes, while C2C shows the opposite behaviour (see Figure~\ref{fig:res}(d,e,i) or Figure~\ref{fig:resG}(i,j,l)). From the user point of view, we believe that it is better to obtain more commissions than omissions. Indeed, as changed parts are rare compared to unchanged ones in general, it is faster to check errors and correct changed predictions than unchanged ones.
 
The point-to-point nearest neighbor correspondence lacks precision in the presence of occlusion in the 3D PCs. Indeed, due to the geometry of acquisition, some occlusions may appear in PCs, these hidden parts may not be similar in the two compared PCs leading to difficulties when comparing points in the DCVA part.

Once again SSL-DCVA seems more interesting than SSST-DCVA when looking at training time. SSST-DCVA takes about 17 hours to train on H3D dataset, while SSL-DCVA only requires about 9 minutes to train (see Table~\ref{tab:res}). The DCVA part on the manually annotated test set takes about 40 seconds. 

Following our experimental assessment, we have identified two difficulties faced by our method. The first one is related to the hypothesis of rare changes required for the temporal consistency loss (Equation~\ref{eq:ltempcons}). Although already used in the literature \citep{saha2022self}, this assumption should be verified considering the training set of the studied dataset, whereas the test set should contain enough changes so that the thresholding operation is valid. For example, this assumption prevents us from applying the presented method on the Urb3DCD dataset \citep{degelis2021change} because it contains a high number of changed objects in the training set. Then, the other issue with our method comes from the DCVA part which relies on a point-to-point comparison based on the nearest point. This point comparison is not optimal in occluded parts as well as in dense urban areas. Let us note that this issue has been already mentioned when describing C2C misclassifications (see region of interest in Figure~\ref{fig:res}(i)). Even if DCVA comparison relies on multiple deep features, avoiding the problem when the two points being compared have different latent embeddings, the problem remains when the latent embedding of the two points under comparison is similar, meaning the same class is predicted by the back-bone network (whether trained by transfer learning or self-supervision). Indeed, in this case, the deep magnitude coefficient computed from deep features will be similar.

Finally, based on the state-of-the-art in unsupervised 2D image change detection, we proposed to adapt the SSL strategy developed in \cite{saha2022self} to 3D PCs change detection task. Our study showed the possibilities offered by SSL to tackle this task. However, there is still room for improvement. Indeed, different SSL strategies have been already developed in the literature for 3D point clouds understanding \citep{sauder2019self,xie2020pointcontrast,alliegro2021joint,chen2021shape, zhang2021self} or 2D image change detection \citep{leenstra2021self, cai2021task, chen2022self, saha2022self}.  SSL is vast, and many different strategies can be elaborated to train a neural network to extract interesting features. Further SSL studies can be conducted to tackle 3D PCs change detection task based on the existing literature in 3D PCs understanding and/or 2D image change detection. Besides, further development of pre-text tasks directly designed for 3D PCs change detection task would be relevant. For example, a possible improvement would be to incorporate a change-related task in the SSL training of the network, so that the learned features are directly related to change. Which will enable us to get rid of the nearest point comparison, one of the possible difficulties of our method.

\section{Conclusion}
\label{sectionConclusion}

In this paper, we have proposed a method able to detect changes into raw 3D PCs using unsupervised deep learning, i.e.,  without any ground truth annotation for the training step. This unsupervised change detection in 3D PCs is challenging due to the lack of point-to-point correspondence between pre-change and post-change 3D points and the proposed method effectively addressed this problem. It further exploits self-supervised learning through deep clustering and contrastive learning to effectively characterize the target area. The method also relies on an adaptation of the deep change vector analysis framework to the particular case of 3D PCs data. Experiments on the public AHN dataset demonstrate both the effectiveness of the proposed approach over existing unsupervised approaches and the additional benefit brought by self-supervised learning w.r.t. transfer learning (i.e., using a network pre-trained on another public dataset). More specifically, our method reach 85\% of mean accuracy. It further allows to increase the best traditional and  unsupervised learning method by around 9\% and 4\% of mean of IoU respectively.  Nevertheless, the performance of the proposed unsupervised method cannot compete with supervised methods yet, that provide very accurate results, at the cost of tedious annotation of large datasets though.
In the future, we will focus on further improving the method in the occluded regions, e.g., by taking inspiration from the recent literature on contrastive learning in the scene boundaries \citep{tang2022contrastive} or to directly make the network learning change-related features. Finally, let us emphasize that our self-supervised method is based on a learning process. As such, it should be agnostic to the sensor used to provide the 3D PCs. Thereby, we consider as future work to experiment our method with other types of 3D data such as photogrammetric or terrestrial LiDAR scanning.

\section*{Acknowledgements}
This work was granted access to the HPC resources of IDRIS under the allocation 2021-AD011011754R1 made by GENCI. The research is also funded by the German Federal Ministry of Education and Research (BMBF) in the framework of the international future AI lab ``AI4EO -- Artificial Intelligence for Earth Observation: Reasoning, Uncertainties, Ethics and Beyond'' (grant number: 01DD20001).

\bibliographystyle{unsrtnat}
\bibliography{references}  






\end{document}